\documentclass[aps,prb,superscriptaddress,reprint,showpacs,floatfix]{revtex4-1}
\usepackage{graphicx}% Include figure files
\usepackage{dcolumn}% Align table columns on decimal point
\usepackage{bm}% bold math
\usepackage{xcolor}
\usepackage{relsize}
\usepackage{amsmath}
\usepackage{amsfonts}
\usepackage{mathtools}
\usepackage{braket}
\usepackage{gensymb}
\usepackage{url}
\usepackage[colorlinks = true,
            linkcolor = blue,
            urlcolor  = blue,
            citecolor = blue,
            anchorcolor = blue]{hyperref}
\usepackage{footmisc}
\usepackage[utf8]{inputenc}
\usepackage{hyperref}
\usepackage[T1]{fontenc}

\begin{document}

\preprint{APS/123-QED}

\title{Optical analogue of Dresselhaus spin-orbit interaction in photonic graphene}

\author{C. E. Whittaker}
\email{charles.whittaker@sheffield.ac.uk}
\affiliation{Department of Physics and Astronomy, University of Sheffield, Sheffield S3 7RH, United Kingdom}%
\author{T. Dowling}
\affiliation{Department of Physics and Astronomy, University of Sheffield, Sheffield S3 7RH, United Kingdom}
\author{A. V. Nalitov}
\affiliation{Faculty of Science and Engineering, University of Wolverhampton, Wulfruna Street, Wolverhampton WV1 1LY, UK}
\affiliation{Science Institute, University of Iceland, Dunhagi-3, IS-107 Reykjavik, Iceland}
\affiliation{ITMO University, St. Petersburg 197101, Russia}%
\author{A. V. Yulin}
\affiliation{ITMO University, St. Petersburg 197101, Russia}%
\author{B. Royall}
\affiliation{Department of Physics and Astronomy, University of Sheffield, Sheffield S3 7RH, United Kingdom}%
\author{E. Clarke}
\affiliation{EPSRC National Epitaxy Facility, University of Sheffield, Sheffield S1 3JD, United Kingdom}
\author{M. S. Skolnick}
\affiliation{Department of Physics and Astronomy, University of Sheffield, Sheffield S3 7RH, United Kingdom}
\affiliation{ITMO University, St. Petersburg 197101, Russia}%
\author{I. A. Shelykh}
\affiliation{Science Institute, University of Iceland, Dunhagi-3, IS-107 Reykjavik, Iceland}
\affiliation{ITMO University, St. Petersburg 197101, Russia}%
\author{D. N. Krizhanovskii}
\email{d.krizhanovskii@sheffield.ac.uk}
\affiliation{Department of Physics and Astronomy, University of Sheffield, Sheffield S3 7RH, United Kingdom}
\affiliation{ITMO University, St. Petersburg 197101, Russia}%

%\date{\today}

\maketitle

\textbf{The concept of gauge fields plays a significant role in many areas of physics from particle physics and cosmology to condensed matter systems, where gauge potentials are a natural consequence of electromagnetic fields acting on charged particles and are of central importance in topological states of matter\cite{RevModPhys.83.1057}. Here, we report on the experimental realization of a synthetic non-Abelian gauge field for photons \cite{Chen2019} in a honeycomb microcavity lattice\cite{PhysRevLett.112.116402}. We show that the %in-plane 
effective magnetic field associated with TE-TM splitting has the symmetry of Dresselhaus spin-orbit interaction around Dirac points in the dispersion, and can be regarded as an SU(2) gauge field \cite{PhysRevLett.114.026803}. The symmetry of the field is revealed in the optical spin Hall effect (OSHE), where under resonant excitation of the Dirac points precession of the photon pseudospin around the field direction leads to the formation of two spin domains. Furthermore, we observe that the Dresselhaus field changes its sign in the same Dirac valley on switching from $s$ to $p$ bands in good agreement with the tight binding modelling. Our work demonstrating a non-Abelian gauge field for light on the microscale paves the way towards manipulation of photons via spin on a chip.}

%This result is in marked contrast to the four domain pattern observed when the excitation is close to the $\Gamma$ point of the Brillouin zone (BZ) and the case of a planar unpatterned cavity.}

Gauge fields are central to the description of fundamental forces and can carry profound physical consequences. In the case of electromagnetism for example, the significance of the magnetic vector potential $\vec{A}$ is revealed by a quantum mechanical phase shift experienced by charged particles in the celebrated Aharonov-Bohm (AB) effect. Whilst this is a manifestation of a U(1) Abelian gauge field with scalar components, there also exist spin-dependent vector potentials with non-commuting components which were first considered by Yang and Mills, i.e. SU(2) non-Abelian gauge fields \cite{PhysRev.96.191}. In condensed matter physics, the non-Abelian framework is highly relevant to the theory of spin-orbit coupling (SOC) in solids \cite{RevModPhys.65.733,Jin_2006}, which plays an indispensable role in the family of spin Hall effects \cite{RevModPhys.87.1213}, topological insulators and superconductors \cite{RevModPhys.83.1057}, and the operation of spintronic devices \cite{Ohno2016}. On the other hand, photons – neutral particles with zero magnetic moment – can also behave as if affected by both Abelian and non-Abelian gauge fields in suitably designed environments \cite{AIDELSBURGER2018394}. These allow the exploration of gauge field Hamiltonians in the optical domain, and a means of manipulating light trajectories and internal degrees of freedom such as spin (polarization) for spinoptronic signal processing applications \cite{doi:10.1002/9783527610150.ch9}. Abelian gauge fields have been engineered in diverse platforms including silica waveguides \cite{Rechtsman2013}, metamaterials \cite{doi:10.1002/adom.201801582}, silicon ring resonators \cite{Hafezi2013,PhysRevLett.113.087403} and liquid crystal optical cavities\cite{PietkaScience}. By contrast, the realization of non-Abelian gauge fields in photonic microstructures enabling manipulation of light via spin on a chip remains a significant challenge.   

%Another approach which has recently emerged is the use of optical microcavities, where the TE-TM splitting (or photonic SOC) can serve as the basis of an artificial gauge field. Whilst alone it does not constitute a gauge field due to its quadratic momentum dependence, in the presence of other terms which modify the dispersion a gauge-field Hamiltonian can be engineered \cite{PhysRevLett.112.066402}. For example, a recent work utilized an optical microcavity containing Perovskite flakes which induce a strong polarization anisotropy, creating regions of the dispersion analogous to that of an electronic system with Rashba SOC \cite{1912.09684}. 

%One of the ways to implement a non-Abelian photonic gauge field on a microscale in monolithic structures is to use the reduced spatial symmetry of laterally patterned microcavities along with the polarization degree of freedom  \cite{PhysRevLett.114.026803,Solnyshkov2016}. 

One possible way to implement artificial non-Abelian gauge fields on the microscale in a monolithic structure is to use the reduced spatial symmetry of laterally patterned semiconductor microcavities along with the native TE-TM splitting (photonic SOC) \cite{PhysRevLett.114.026803,Solnyshkov2016}. Honeycomb lattices are of particular interest, since they provide access to the physics of graphene and related materials, including the Dirac dispersion \cite{PhysRevLett.112.116402}, edge states \cite{Mili_evi__2015} and influence of strain \cite{PhysRevX.9.031010}, all in a controlled photonic environment in which some of the limitations of real graphene can be overcome. Importantly, whilst  graphene itself suffers from small SOC which prevents observation of the spin Hall effect, the photonic SOC can be enhanced in wavelength-scale photonic lattices \cite{Sala2015,PhysRevLett.120.097401} enabling the physics of non-Abelian gauge fields in graphene geometries to be explored. 

\begin{figure}
\centering
\includegraphics[scale=1]{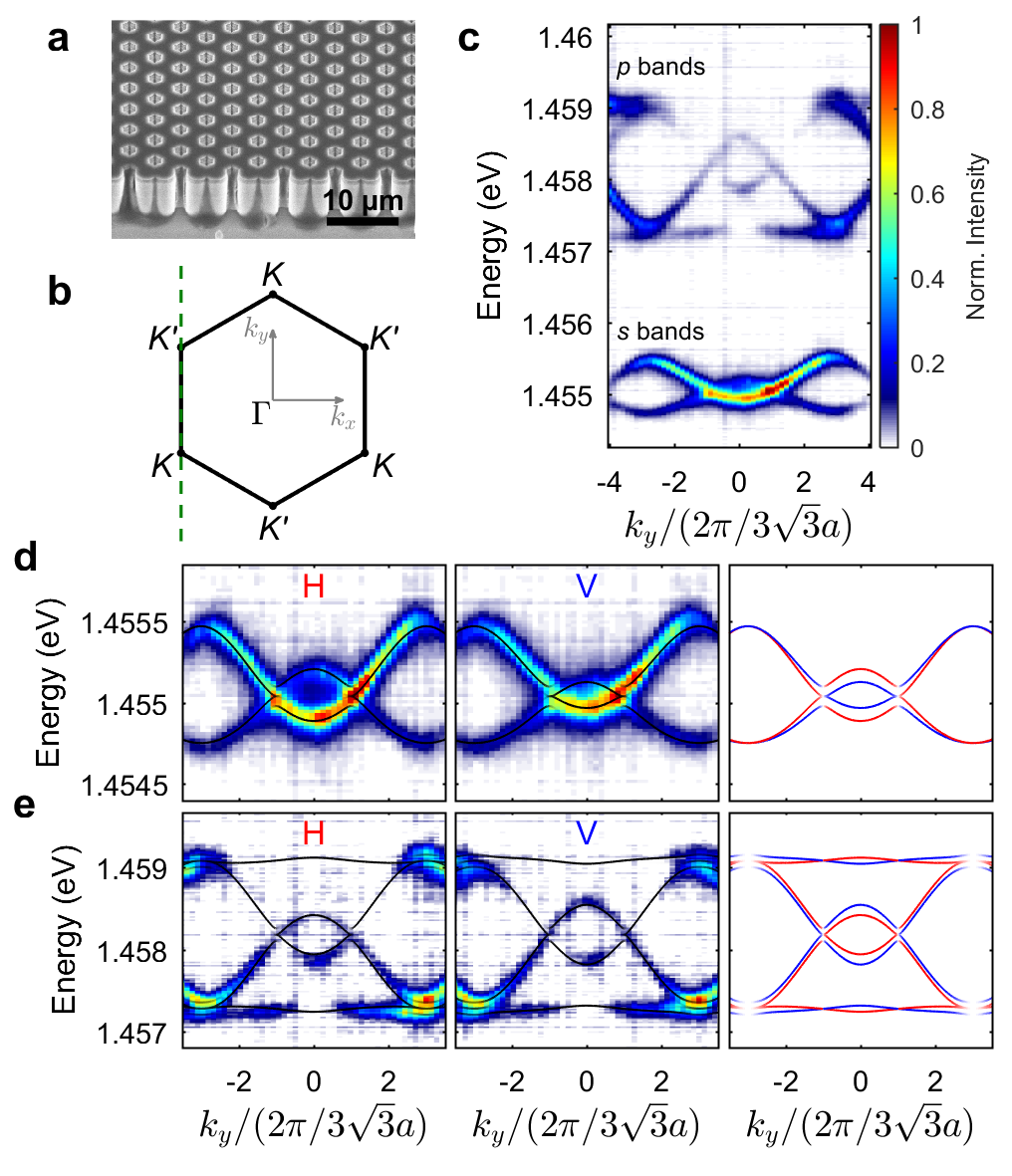}
\caption{\textbf{Photonic graphene sample and dispersion relations.} \textbf{a}, Scanning electron microscope image of the honeycomb lattice. \textbf{b}, Schematic of the first BZ. \textbf{c}, Angle-resolved PL spectrum along the dashed green line in (\textbf{b}). \textbf{d}, Polarization-resolved emission of $s$ bands along the direction shown in (\textbf{c}) with tight binding calculations (solid curves). Left, H polarization. Middle, V polarization. Right, theoretical $S_1$ Stokes parameter. \textbf{e}, Same as (\textbf{d}) for the $p$ bands. $a$ denotes the separation of 2.8 $\mu$m between adjacent pillars. The red and blue in (\textbf{d}) and (\textbf{e}) correspond to H and V polarizations respectively.
}
\label{fig1}
\end{figure}

In this Letter we utilize a patterned GaAs-based microcavity with a honeycomb lattice geometry (Fig. \ref{fig1}a), i.e. photonic graphene, to study the influence of photonic SOC on the dispersion. In this setting (see \hyperref[sec:methods]{Methods} for sample details), which was previously considered theoretically in ref. \onlinecite{PhysRevLett.114.026803}, the interplay between the SOC and the reduced spatial symmetry imposed by the lattice transforms the double winding effective magnetic field associated with TE-TM splitting into a Dresselhaus-type field with a single winding locally around the Dirac points, which can be described in terms of a non-Abelian gauge field. Here, we visualize the field texture around these high symmetry points, and further confirm the Dresselhaus symmetry by the OSHE revealing the formation of two cross-polarized spin domains.
%by measuring the angle- and polarization-resolved PL from the lattice across momentum space. We then perform measurements of the OSHE under resonant excitation of the Dirac points, confirming the Dresselhaus symmetry of the field.
%Furthermore, we show that the effect can also be observed in the higher energy $p$ bands. 
Our results are in good agreement with the theory presented in ref. \onlinecite{PhysRevLett.114.026803} and demonstrate the potential for the engineering of artificial gauge fields for photons in different orbital bands using model 2D lattice systems.  Finally, we note that recently the non-Abelian AB effect has been observed by cascading non-reciprocal optical elements in a fiber-optic setup (i.e not on a microscale and governed by physical mechanisms very different from those reported in the present manuscript) \cite{Yang1021}.

To measure the dispersion relation of our sample we use low-power incoherent excitation to populate all modes. The band structure features linear Dirac crossings at characteristic momenta, namely the $K$ and $K'$ points at the Brillouin zone (BZ) corners (Fig. \ref{fig1}b), which are visible in the angle-resolved PL spectra in Fig. \ref{fig1}c, where both the fundamental $s$ bands and higher energy $p$ bands are displayed. By resolving the emission in linear polarization, the orientation of the TE-TM splitting effective magnetic field at each energy and momentum can be revealed since it corresponds to the pseudospin of the eigenstate. % \cite{Shelykh_2009}. 
Hence, in order to characterize the pseudospin texture and therefore the field orientation across momentum space, we measure the first two Stokes parameters $S_1$ and $S_2$ of the emission (see \hyperref[sec:methods]{Methods}). For both $s$ and $p$ bands, shown in Fig. \ref{fig1}d,e respectively, a pronounced splitting between TE and TM modes (which have horizontal (H) and vertical (V) polarization for the direction plotted) is visible, which is well described using a tight binding formalism including SOC as in refs. \onlinecite{PhysRevLett.114.026803,Zhang_2019}. Importantly, if one follows the innermost bands it can be seen that the pseudospin changes sign when passing through the $K$ and $K'$ points ($k_y/(2\pi/3\sqrt{3}a)=\pm1$). In order to study this polarization behaviour in more detail we construct 2D energy-resolved polarization maps using tomographic imaging (see \hyperref[sec:methods]{Methods}). 

\begin{figure*}
\centering
\includegraphics[scale=1]{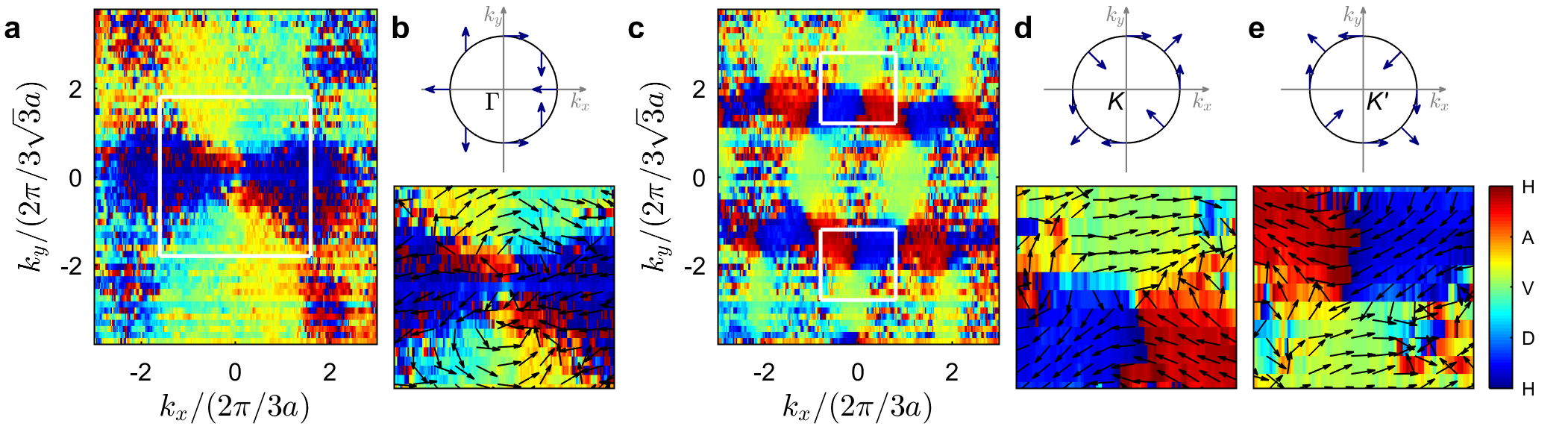}
\caption{\textbf{Texture of the effective magnetic fields surrounding $\Gamma$, $K$ and $K'$ points.} \textbf{a}, Momentum space map of $\phi$ at the energy of the $\Gamma$ point (1.4547 eV). \textbf{b}, Close-up of region delimited by the white rectangle in \textbf{a}, with arrows representing the pseudospin vector. The winding of the field around the $\Gamma$ point for fixed magnitude of the wave vector is shown schematically on top. \textbf{c}, Momentum space map of $\phi$ at the energy of the Dirac points (1.4551 eV). \textbf{d}, \textbf{e}, Close-up of regions delimited by the upper and lower white rectangles in \textbf{c}, corresponding to $K$ and $K'$ points respectively, with arrows representing the pseudospin vector. The winding of the field around the $K$ and $K'$ points for fixed magnitude of the wave vector is shown schematically on top. D and A denote diagonal and anti-diagonal polarizations.}
\label{fig2}
\end{figure*} 

First we will focus on the $s$ bands. In Fig. \ref{fig2} we see 2D maps of the linear polarization angle $\phi$ in momentum space, calculated as $2\phi = \arctan{(S_{2}/S_{1})}$. For the $\Gamma$ point, which corresponds to the emission around $\vec{k}=0$ at the energy minimum of the dispersion, a quadrupole pattern can be seen (Fig. \ref{fig2}a). A close-up of the region around the $\Gamma$ point is plotted overlaid with arrows showing the pseudospin texture, which reveals the orientation of the effective magnetic field, in Fig. \ref{fig2}b. It has the familiar form of the dipolar field associated with TE-TM splitting in microcavities. %\cite{SOLNYSHKOV2016920}.
%, as depicted by the schematic showing the double winding azimuthal dependence. 
Fig. \ref{fig2}c shows the corresponding map at the energy of the Dirac points. The points around which the field winds are now the $K$ and $K'$ points at the six vertices of the hexagonal BZ(s). However, in contrast to the $\Gamma$ point, the local symmetry no longer has a double azimuthal dependence. 
The pseudospin texture reveals a single winding of the field with the characteristic texture of a Dresselhaus field \cite{Dresselhaus1955} as shown in close-ups of the $K$ and $K'$ points (Fig. \ref{fig2}d,e). We note that the local effective magnetic fields have opposite sign around the $K$ and $K'$ points, although the direction of field rotation is the same (counter-rotating with the azimuthal angle) in agreement with theory \cite{PhysRevLett.114.026803}.
%Fig. \ref{fig2}d,e show close-ups of the $K$ and $K'$ points found at $k_{y}/(2\pi/3\sqrt{3}a) = \pm 2$ respectively, where the pseudospin texture reveals a single winding of the field with the characteristic texture of a Dresselhaus field \cite{Dresselhaus1955}, as depicted by the schematics. 
%The pseudospin patterns around the Dirac points reveal the direction of an effective magnetic field sharing the same angular dependence as that due to the Dresselhaus spin-orbit coupling \cite{Dresselhaus1955}. 
As we show in the \hyperref[sec:methods]{Methods}, using a suitable minimal coupling transformation the observed effective field around the Dirac points can be described in terms of a non-Abelian gauge field with non-commuting components. %Furthermore, the local effective magnetic fields have opposite sign around the $K$ and $K'$ points, although the direction of field rotation is the same (counter-rotating with the azimuthal angle) in agreement with our theory. 
%The reversal of the sign of the effective magnetic field with valley index is in agremeent with our theory. 
By contrast, around the $\Gamma$ point the SOC has the same form as for unpatterned microcavities and cannot be described in terms of a synthetic gauge field.

\begin{figure}
\centering
\includegraphics[scale=1]{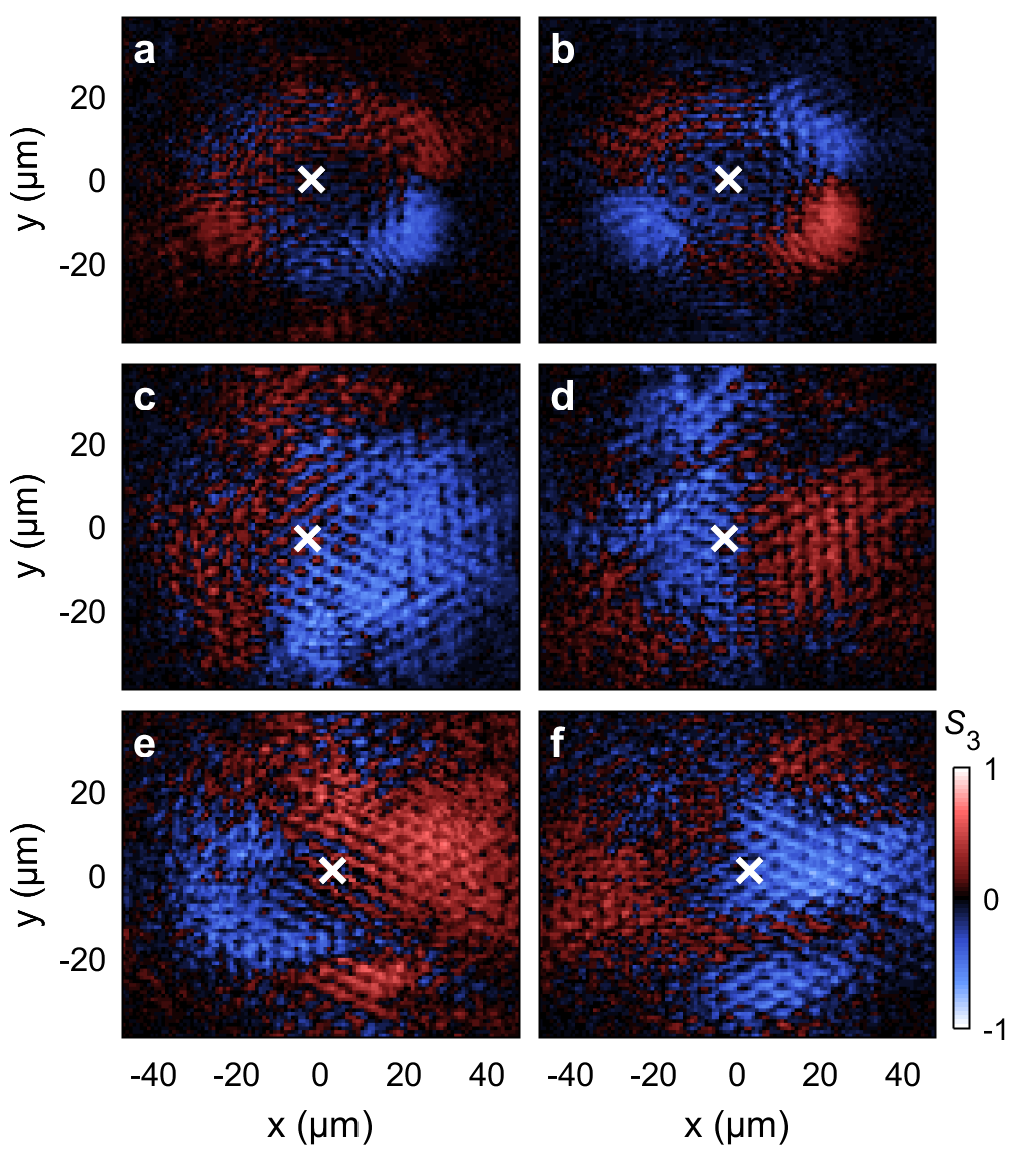}
\caption{\textbf{Observation of the optical spin Hall effect.} \textbf{a}, \textbf{c}, \textbf{e}, Measured real space circular polarization degree $S_3$ under resonant excitation at the $\Gamma$ (\textbf{a}), $K$ (\textbf{c}) and $K'$ (\textbf{e}) points with a H polarized pump. \textbf{b}, \textbf{d}, \textbf{f}, Corresponding results for a V polarized pump. The energies used for excitation of the $\Gamma$ and $K/K'$ points are 1.4547 eV and 1.4551 eV respectively. The cross in each panel marks the position of the pump spot.}
\label{fig3}
\end{figure} 

One of the clearest manifestations of the effective magnetic field acting on photons
%associated with TE-TM splitting in microcavities 
is the formation of spin currents in the OSHE, caused by pseudospin precession around the $k$-dependent effective magnetic field \cite{PhysRevLett.95.136601}. At a given energy, the wave vector and polarization of the injected light can be used to control the spin texture of the emission since the pseudospin rotation depends on the relative angle between its initial direction and the effective field. We use this knowledge to unveil the different symmetries shown in Fig. \ref{fig2} by imaging the time-integrated real space emission of our sample under continuous optical excitation at the $\Gamma$, $K$ and $K'$ points with a linearly polarized pump. We vary the energy and angle of the incoming laser to excite these points in the dispersion and measure the resulting emission intensity in right and left circular polarizations to determine the Stokes parameter $S_3$ (see \hyperref[sec:methods]{Methods}).
In Fig. \ref{fig3}a we show how resonant excitation at the $\Gamma$ point under H polarized excitation indeed leads to the observation of four domains with alternating circular polarization, confirming that at the energy minimum of the dispersion, at the centre of the BZ, the effective magnetic field in Fig. \ref{fig2}b has the same form as that of conventional planar microcavities \cite{Leyder2007, PhysRevLett.114.026803}.
%In the conventional case of unpatterned planar microcavities, the signature of the OSHE is the four alternating circularly polarized domains which appear in real and momentum space \cite{Leyder2007,PhysRevLett.109.036404} as a result of the double winding texture of the TE-TM field. Since we observe this same dipolar symmetry around the $\Gamma$ point in momentum space (Fig. \ref{fig2}b), we would expect the same pattern in real space under resonant excitation in the OSHE regime \cite{PhysRevLett.114.026803}. 
%In Fig. \ref{fig3}a we show how resonant excitation at this point under H polarized excitation indeed leads to the observation of four domains with alternating circular polarization. This confirms that at the energy minimum of the dispersion, at the centre of the BZ, the effective magnetic field has the same form as that of conventional planar microcavities.
In contrast, under H polarized pumping at the $K$ point (Fig. \ref{fig3}c), only two domains are seen, formed to the left and right of the pump spot and with opposite circular polarizations as expected from the Dresselhaus symmetry surrounding the Dirac points (Fig. \ref{fig2}d,e).
%the Dresselhaus symmetry surrounding the Dirac points (Fig. %\ref{fig2}d,e) should instead give rise to two domains since the %field has only a single winding. Indeed, under H polarized pumping %at the $K$ point (Fig. \ref{fig3}c), only two domains are seen, %formed to the left and right of the pump spot and with opposite %circular polarizations.
This is expected since the injected pseudospin vector lies parallel/anti-parallel to the field direction along the $k_y$ axis in the locally excited region of momentum space, so there should be no evolution of the pseudospin along $y$. When the excitation angle is changed to excite the $K'$ point instead (Fig. \ref{fig3}e), the pattern is reversed as expected since the sign of the Dresselhaus field is opposite. In Fig. \ref{fig3}b,d,f we demonstrate that upon changing to V polarized excitation, the patterns shown in Fig. \ref{fig3}a,c,e are all reversed since the initial pseudospin vector points in the opposite direction \cite{PhysRevLett.95.136601}, which confirms that the observed spin patterns result from precession of the pseudospin vector in the OSHE regime. 

\begin{figure}
\centering
\includegraphics[scale=1]{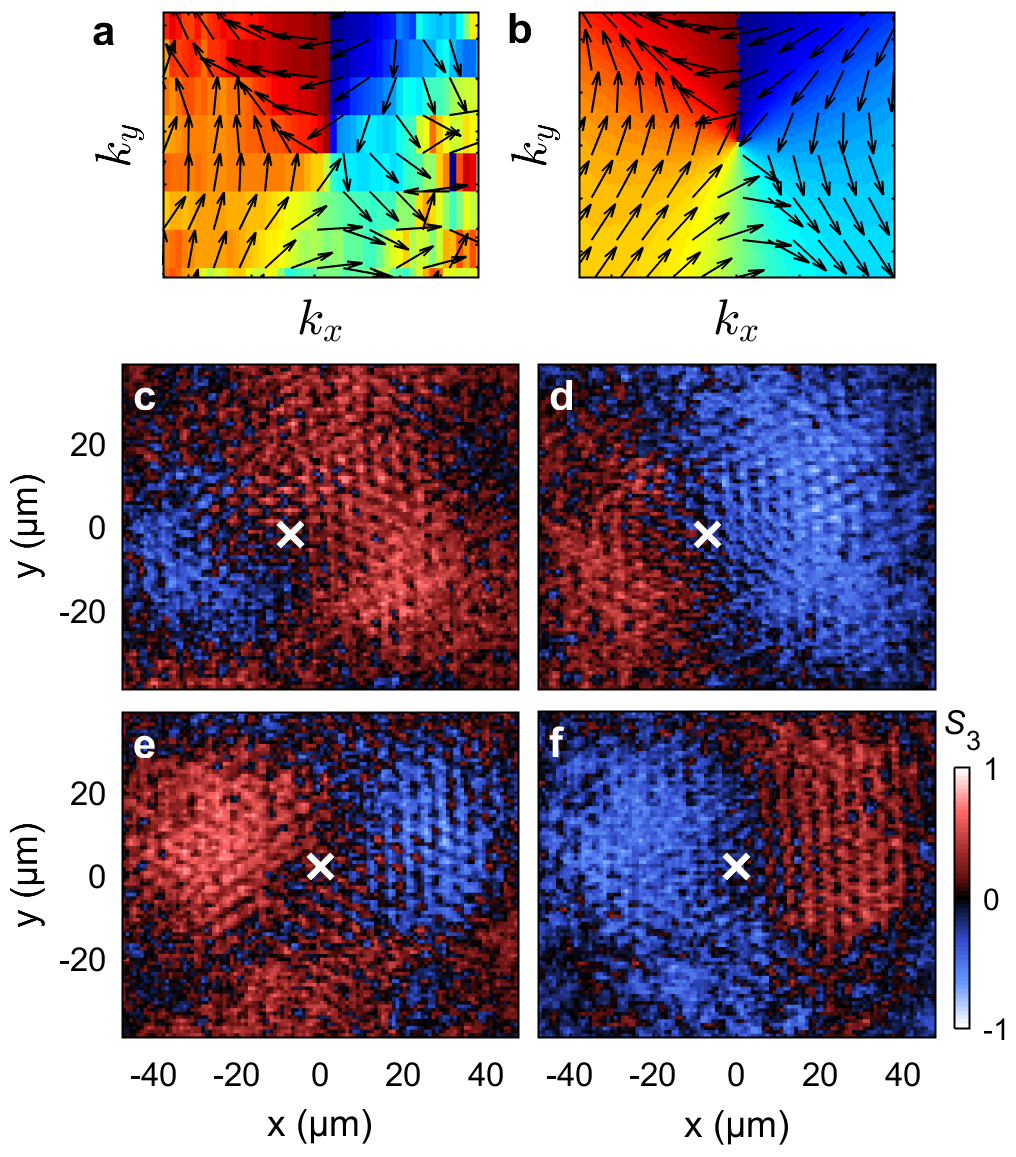}
\caption{\textbf{Effective magnetic field texture and optical spin Hall effect for $p$ bands.} \textbf{a}, Experimentally obtained effective magnetic field texture surrounding a $K$ point. \textbf{b} Corresponding calculated effective magnetic field texture surrounding a $K$ point. \textbf{c}, \textbf{d}, Measured real space circular polarization degree $S_3$ under resonant excitation at the $K$ (\textbf{c}) and $K'$ (\textbf{d}) points with a H polarized pump. \textbf{e}, \textbf{f},  Corresponding results for a V polarized pump. The energy used for excitation is 1.4582 eV. The cross in panels (\textbf{c--f}) marks the position of the pump spot.}
\label{fig4}
\end{figure} 

Now we turn our attention to the $p$ bands. As is the case for the $s$ bands, a reversal of the pseudospin either side of the Dirac points can be seen in Fig. \ref{fig1}e. Using the same procedure as that of Fig. \ref{fig2}, we determine the texture of the effective magnetic field across momentum space at the energy of the Dirac points to reveal the local symmetry surrounding the $K$ and $K'$ points. The full momentum space linear polarization map (see Supplementary information) confirms that the local symmetry around these points has the form of Dresselhaus SOC (with opposite sign for $K$ and $K'$) as is the case for the $s$ bands. We show the field surrounding a $K$ point in Fig. \ref{fig4}a, where a clear single winding of the Dresselhaus type is visible.  Note that the sign of the field for a given valley ($K$ or $K'$) is opposite to the case of the $s$ bands. Our finding is supported by the tight binding model developed for the $p$ orbitals \cite{Zhang_2019}, where the calculated field texture shows excellent agreement (Fig. \ref{fig4}b). To further confirm the Dresselhaus fields, we perform OSHE measurements by coherently exciting the $p$ bands. In Fig. \ref{fig4}c--f we show results for resonant excitation of the $K$ and $K'$ points at $k_{y}/(2\pi/3\sqrt{3}a) = \pm 2$ (Fig. \ref{fig4}c,e respectively). Clear twofold circular polarization patterns can be seen, which rotate when the excitation polarization is changed from H to V (Fig. \ref{fig4}d,f).

In summary, we have experimentally demonstrated the existence of local Dresselhaus fields surrounding the Dirac points in photonic graphene, confirmed by the generation of twofold circular polarization patterns in the optical spin Hall effect. Our findings constitute the realization of a synthetic SU(2) non-Abelian gauge field induced by the presence of the honeycomb periodic potential, which leads to a TE-TM effective magnetic field with a modified texture at specific points in momentum space. We note that whilst such fields may be engineered in other lattices featuring Dirac cones such as Kagome lattices \cite{PhysRevB.94.115437}, it is not possible in other geometries such as Lieb lattices due to the square symmetry. Practically speaking, our findings offer a means of separating and routing spins, where the single winding effective magnetic field (odd in $k$) analogous to electronic systems is highly advantageous since it leads to counter-propagation of opposite spins. In microcavities with a small exciton-photon detuning, the addition of spin-anisotropic polariton-polariton interactions to the present system opens up new possibilities, including spin-dependent Klein tunneling \cite{Solnyshkov2016}, interaction-induced topological phase transitions \cite{PhysRevB.93.085438} and potentially a nonlinear modification of the spin domains \cite{PhysRevLett.110.016404}. 

\section*{Methods}
\label{sec:methods}
\subsection*{Sample description}

Our sample is a GaAs microcavity with 23 (26) top (bottom) GaAs/Al$_{0.85}$Ga$_{0.15}$As distributed Bragg reflector pairs and six In$_{0.04}$Ga$_{0.96}$As quantum wells, which was previously described in ref. \onlinecite{PhysRevB.99.081402}. The sample was processed using electron beam lithography and plasma dry etching to pattern arrays of micropillars with 3 $\mu$m diameters and an etch depth on the order of 8 $\mu$m. We study a honeycomb lattice with a pillar-to-pillar separation of 2.8 $\mu$m, whose size is $\sim$120$\times$100 $\mu$m$^{2}$. 

\subsection*{Excitation scheme}

To characterize the dispersion relation of the honeycomb lattice a low-power non-resonant diode laser is used in reflection geometry to incoherently populate all of the lattice modes. To study the formation of pseudospin domains we excite the lattice in transmission geometry with a continuous wave Ti:sapphire laser tuned to be resonant with the honeycomb lattice dispersion. Different states can be excited by varying the energy and angle of incidence $\theta$ of the Gaussian laser beam, the latter of which is changed using a translation stage to move the lateral position of the beam before the excitation objective. This allows us to accurately control the in-plane wave vector of the injected wave packet since $k$ = $\left(\frac{\omega}{c}\right)\sin{\theta}$ where $\omega$ is the laser frequency. The linear polarization of the injected wave packet can also be controlled through the use of a linear polarizer and half wave plate in the excitation path. The excitation beam has a full width at half maximum of $\sim$15 $\mu$m.     

\subsection*{Detection scheme}

In order to characterize the pseudospin texture around the $\Gamma$, K and K’ points, a half wave plate and linear polarizer are used in the detection path during the non-resonant excitation measurements in order to measure the first two Stokes parameters. These are given by $S_{1} = (I_{H}-I_{V})/(I_{H}+I_{V})$ and $S_{2} = (I_{D}-I_{A})/(I_{D}+I_{A})$ where $I_H$, $I_V$, $I_D$ and $I_A$ give the intensity of emitted light in horizontal, vertical, diagonal and anti-diagonal polarizations respectively. By scanning the final lens across the spectrometer slit, multiple $E$ vs $k$ slices are recorded (corresponding to different wave vectors in the direction orthogonal to the spectrometer slit) allowing 2D energy-resolved polarization maps to be constructed. For the resonant transmission measurements the half wave plate is replaced by a quarter wave plate to measure the third Stokes parameter, which is given by $S_{3} = (I_{\sigma^{+}}-I_{\sigma^{-}})/(I_{\sigma^{+}}+I_{\sigma^{-}})$ where $I_{\sigma^{+}}$ and $I_{\sigma^{-}}$ correspond to the emission intensity in right and left circular polarizations respectively.

\subsection*{Gauge field representation}

The pseudospin patterns in Fig \ref{fig2}c-e correspond to the eigenstates of the polariton graphene effective Hamiltonian, which has the following form in the vicinity of the Dirac points \cite{PhysRevLett.114.026803}:
\begin{equation} \label{eq_H}
H^D(\vec{q}) = \hbar v_F \left( \tau_z q_x \sigma_x + q_y \sigma_y \right) + \Delta \left( \tau_z \sigma_y s_y - \sigma_x s_x \right),
\end{equation}
where $\vec{q}$ is the wave vector deviation from one of the two Dirac points, set by the valley index $\tau_z = \pm 1$, $v_F$ is the effective Fermi velocity, $\vec{\sigma}$ and $\vec{s}$ are the sublattice and the polarization pseudospin operators, and $\Delta$ is the effective photonic SOC strength \cite{Sala2015}.
Prefactors of both terms in Hamiltonian \eqref{eq_H} may be expressed in the tight binding model parameters: $\hbar v_F = 3Ja/2$, $\Delta = 3 \delta J/2$ (see next section in \hyperref[sec:methods]{Methods} for additional details).
The spin-orbit term may be included in the low energy graphene Hamiltonian, represented by the first term in Eq. \eqref{eq_H}, with minimal coupling transformation $\vec{q} \rightarrow \vec{q} - \vec{A}$ with the gauge field components given by

\begin{equation} \label{eq_A}
    A_x = - {\Delta  \over \hbar v_F} \tau_z s_x, \; A_y = {\Delta  \over \hbar v_F}\tau_z s_y.
\end{equation}

The artificial gauge field \eqref{eq_A} is non-Abelian since the components do not commute \cite{Solnyshkov2016}. This field is responsible for the suppression of Klein tunneling \cite{Solnyshkov2016} and emergence of topologically nontrivial band gaps of the polariton spectrum in the presence of external magnetic fields \cite{PhysRevLett.114.116401,Klembt2018}.
Polarization spectral splitting may be also attributed to the effective magnetic field, acting on polariton pseudospin. In the range of energies $\Delta \ll \vert E \vert \ll \hbar v_F / a$ this effective SOC is given by the Hamiltonian term $ H_{\mathrm{SOC}}^D(\vec{q}) = \pm \Delta \left( q_x s_x - q_y s_y \right) / q$, sharing the same angular dependence with the Dresselhaus spin-orbit term \cite{Dresselhaus1955}, but constant in the wave vector absolute value $q$.
Note that the sign of the splitting inverts with both valley index $\tau_z$ and the sign corresponding to upper and lower Dirac cones.

The Hamiltonian \eqref{eq_H} close to Dirac points thus drastically differs from its counterpart in the vicinity of the $\Gamma$ point 
\begin{equation} \label{eq_G}
H^\Gamma(\vec{k}) = \hbar^2 k^2 / (2m)+\beta\left[s_x(k_x^2 - k_y^2)+2s_yk_x k_y\right],
\end{equation}
where the first term corresponds to a free particle with the effective mass $m = \hbar^2/(3 J a^2)$ and the second term describes the action of TE-TM splitting, corresponding to the effective magnetic field with components $\Omega_x = \beta a^2 (k_x^2 - k_y^2)$, $\Omega_y = 2 \beta k_x k_y$ with $\beta = 3 \delta J a^2/8$, related to the tight binding model parameters (see next section in \hyperref[sec:methods]{Methods}  for details). Note that the quadratic dependence of the effective magnetic field on the components of $\vec{k}$, fully similar to those reported for the case of an unpatterned cavity, precludes its description in terms of the minimal coupling to a synthetic gauge field and leads to the difference of the effective masses of the longitudinal and transverse polariton modes, $m_{l,t}=m(1\pm 2m\beta/\hbar^2)$ \cite{PhysRevB.98.155428}.

\subsection*{Effective field derivation}

The effective magnetic field acting on polariton pseudospin is obtained by development of the tight binding Hamiltonian (see ref. \onlinecite{PhysRevLett.114.026803}):
\begin{equation} \label{eq_H_full}
    H_\mathbf{k} = -J\sigma_+ - \delta J \sigma_+ \otimes \left( f_\mathbf{k}^+ s_+ + f_\mathbf{k}^- s_- \right) + \textrm{H.c.}
\end{equation}
While the local gauge fields in the vicinities of Dirac points K and K$^\prime$ were studied in ref. \onlinecite{PhysRevLett.114.026803}, the effective field given by formula \eqref{eq_G}  near the $\Gamma$ point, reproducing the symmetry and quadratic $k$ dependence of the TE-TM field in planar microcavities, is also inherent to the Hamiltonian \eqref{eq_H_full}.
In the latter case, the splitting is due to the second order terms in the complex coefficients
\begin{equation}
    f_\mathbf{k} = 3\left( 1 - {k^2a^2\over 4} \right), \; f_\mathbf{k}^\pm = - {3 i \over 2} k_\pm a -{3 \over 8} k_\pm^2 a^2,
\end{equation}
where $k_\pm = k_x \pm i k_y$.
The energy dispersion near the ground state in the corresponding order then reads
\begin{equation}
    E_k^\pm  = -3J\left[ 1 - {k^2 a ^2 \over 8} \left( 1- {\delta J^2 \over J^2} \pm {\delta J \over J}  \right) \right],
\end{equation}
and the Hamiltonian has the form of interaction with the effective TE-TM field \eqref{eq_G} of strength given by $\beta= 3 \delta J a^2/8$.

\section*{Data availability}

The data that support the findings of this study are
openly available from the University of Sheffield repository.

\bibliographystyle{ieeetr}
\bibliography{References}

\section*{Acknowledgements}

The work was supported by EPSRC Grants EP/N031776/1 and EP/R04385X/1, and the Government of the Russian Federation through the Megagrant 14.Y26.31.0015. Theoretical modelling performed by A.V.N was supported by Russian Science Foundation Project No. 18-72-10110.

\section*{Author contributions}
CEW and TD performed the experiments and analyzed the data; EC grew the sample; BR performed post-growth fabrication; AVN, AVY and IAS provided theoretical input; CEW and DNK designed the experiment; CEW, AVN, IAS, MSS and DNK wrote the manuscript.

\section*{Competing interests}
The authors declare no competing interests.

%%%%%%%%%% Merge with supplemental materials %%%%%%%%%%
\clearpage
\widetext
\begin{center}
\textbf{\large Supplementary material}
\end{center}
%%%%%%%%%% Merge with supplemental materials %%%%%%%%%%
%%%%%%%%%% Prefix a "S" to all equations, figures, tables and reset the counter %%%%%%%%%%
\setcounter{equation}{0}
\setcounter{figure}{0}
\setcounter{table}{0}
\setcounter{page}{1}
\makeatletter
\renewcommand{\theequation}{S\arabic{equation}}
\renewcommand{\thefigure}{S\arabic{figure}}
%\renewcommand{\bibnumfmt}[1]{[S#1]}
%\renewcommand{\citenumfont}[1]{S#1}
%%%%%%%%%% Prefix a "S" to all equations, figures, tables and reset the counter %%%%%%%%%%

\section{Experimental determination of effective magnetic field textures}

\begin{figure*}[h!]
\centering
\includegraphics[width=0.5\textwidth]{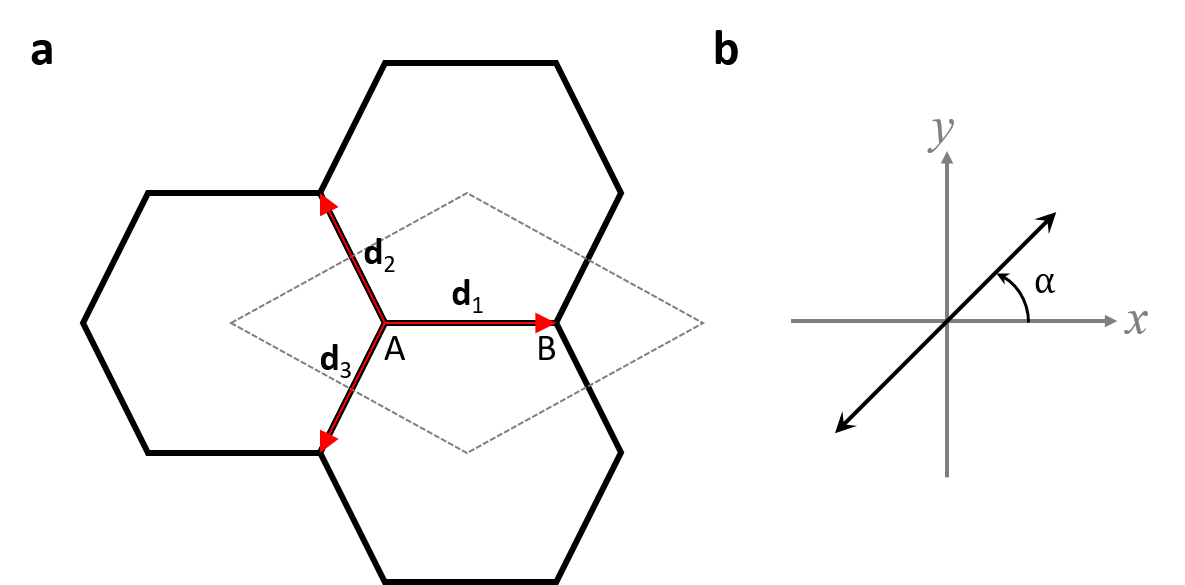}
\caption{(a) Real space orientation of the honeycomb lattice, labelled with $A$ and $B$ sublattices and nearest neighbour vectors $\mathbf{d}_1$, $\mathbf{d}_2$ and $\mathbf{d}_3$. The dashed diamond delimits a unit cell. (b) Definition of linear polarization detection angle $\alpha$. }
\label{schema}
\end{figure*}

In order to extract the effective magnetic field textures in momentum space, we resolve the emission in horizontal (H), diagonal (D), vertical (V) and anti-diagonal (A) polarizations corresponding to detection angles of $0\degree$, $45\degree$, $90\degree$ and $135\degree$ respectively. The detection angle, which we denote as $\alpha$, is defined with respect to the $x$ axis, shown along with the real space orientation of the honeycomb lattice in Fig. \ref{schema}. Tomographic measurements in the H-V and D-A bases allow us to construct the Stokes parameters $S_1=(I_H-I_V)/(I_H+I_V)$ and $S_2=(I_D-I_A)/(I_D+I_A)$ respectively, allowing a complete characterization of the linear polarization state of the emission across momentum space. This is sufficient to determine the effective magnetic field texture since it is entirely in-plane (it has no circular component) in the absence of a real Zeeman field.

\begin{figure*}[h!]
\centering
\includegraphics[width=0.9\textwidth]{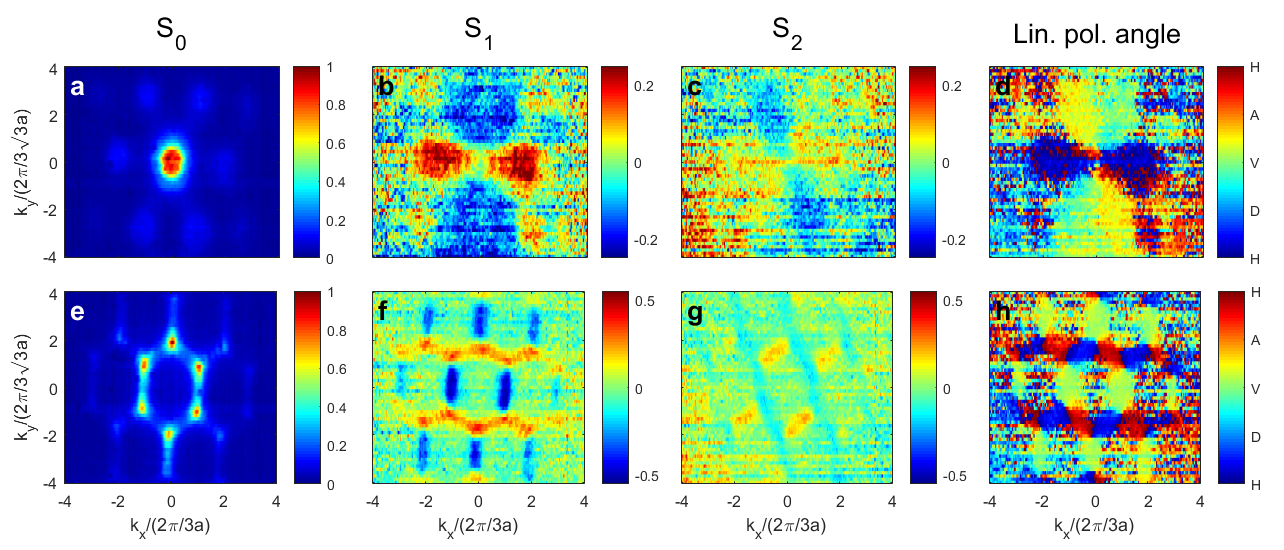}
\caption{(a--d) Tomographic images of the momentum space emission at the energy of the $\Gamma$ point, showing the Stokes parameters $S_0$ (a), $S_1$ (b) and $S_2$ (c), along with the linear polarization angle $\phi$ (d). (e--h) Tomographic images of the momentum space emission at the energy of the Dirac points, showing the Stokes parameters $S_0$ (e), $S_1$ (f) and $S_2$ (g), along with the linear polarization angle $\phi$ (h).}
\label{smaps}
\end{figure*} 

In Fig. \ref{smaps} we show the polarization-resolved momentum space emission in the $s$ bands (corresponding to Fig. 2 of the main text), at the energy of the $\Gamma$ point at 1.4547 eV (a--d) and the Dirac points at 1.4551 eV (e--h). The total intensity $S_0$ reveals emission from the centre of the Brillouin zone (BZ) at $k=0$ in the former case and from the six corners of the BZ forming a hexagon in the latter case. Replicas in the second BZs can also be seen. The $S_1$ and $S_2$ maps shown allow the linear polarization angle $\phi$ of the emission to be calculated at each point using $2\phi=\arctan{(S_2/S_1)}$. Note the factor of two which represents the fact that one only has to rotate a polarization by $180\degree$ in real space to return to the same polarization, whilst the Stokes vector has undergone a full $360\degree$ rotation. 

\begin{figure*}[h!]
\centering
\includegraphics[width=0.9\textwidth]{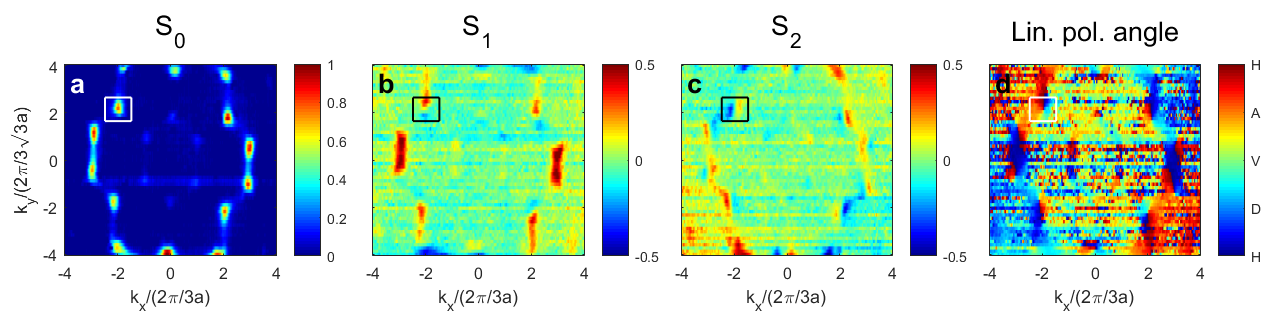}
\caption{(a--d) Tomographic images of the momentum space emission at the energy of the Dirac points in the $p$ bands, showing the Stokes parameters $S_0$ (a), $S_1$ (b) and $S_2$ (c), along with the linear polarization angle $\phi$ (d).}
\label{pmaps}
\end{figure*} 

In Fig. \ref{pmaps} we show the corresponding momentum space maps for the energy of the Dirac points in the $p$ bands. Since the photoluminescence (PL) intensity is much weaker in the $p$ bands, as can be seen in Fig. 1c of the main text, the emission is integrated over a small spectral window of $\sim0.2$ meV between 1.4581 eV and 1.4583 eV to increase the signal. From the total intensity $S_0$ we note that the emission pattern is the same as for the energy of the Dirac points in the $s$ bands, with the difference that the brightest points are no longer found at the corners of the first BZ but at the outer corners of the second BZs. This reflects the larger leakage emission intensity (probably of escaping the cavity) for higher $k$ vectors at higher energy. The rectangle in each panel corresponds to the $K$ point for which the linear polarization angle is shown in Fig. 4 of the main text, which is taken from the second BZ due to the higher PL intensity. 

\section{Resonant excitation}

To resonantly excite specific states in the dispersion relation of our honeycomb lattice we control both the energy and incoming angle of the excitation laser. The excitation energy and $k$ vector corresponding to the results shown in Fig. 3 and 4 of the main text are shown in relation to the band structure by black markers in Fig. S4. The three lower energy points correspond to $s$ band excitation [Fig. 3] and the two upper energy points correspond to $p$ band excitation [Fig. 4]. The $\Gamma$, $K$ and $K'$ points are excited with angles of $0\degree$, $+6.7\degree$ and $-6.7\degree$ respectively.  

\begin{figure*}[h!]
\centering
\includegraphics[width=0.36\textwidth]{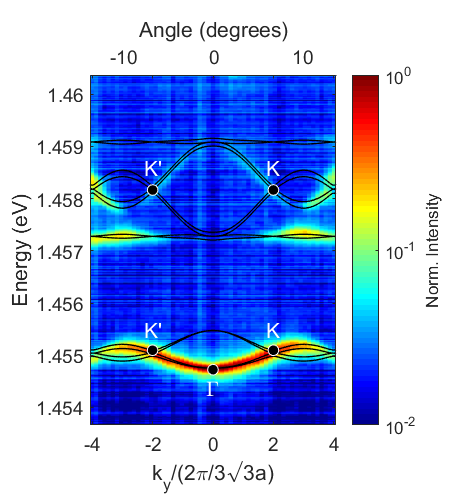}
\caption{Dispersion relation measured under low-power non-resonant excitation along $k_x = 0$ ($K'-\Gamma-K$ direction). The black dots show the excitation energy and $k$ vector corresponding to the results shown in Figs. 3 and 4 of the main text. The solid curves show the band structure calculated using the developed tight binding models.}

\end{figure*} 

% \section{Experimental real space images}

% In our resonant excitation experiments the real space emission is resolved in the circular polarization ($\sigma^{+}$/$\sigma^{-}$) basis in order to determine $S_3=(I_{\sigma^{+}}-I_{\sigma^{-}})/(I_{\sigma^{+}}+I_{\sigma^{-}})$. In Fig. S5 we show the total intensity ($S_{0} = I_{\sigma^{+}} + I_{\sigma^{-}}$) of the emission corresponding to Fig. 3 of the main text. The black cross marks the position of the pump spot and the white box delimits the regions shown in Fig. 3 of the main text. %A small sample tilt of $\sim2\degree$ is visible, which does not affect the findings of our paper.  

% \begin{figure*}[h!]
% \centering
% \includegraphics[width=0.5\textwidth]{rsp_S0.png}
% \caption{(a), (c), (e) Measured real space total intensity $S_0$ under resonant $s$ band excitation at the $\Gamma$ (a), $K$ (c) and $K'$ (e) points with a horizontally polarized pump. (b), (d), (f) Corresponding results for a vertically polarized pump.}
% \end{figure*} 

% In Fig. S6 we show the total intensity data for resonant excitation of the $p$ bands, corresponding to Fig. 4 of the main text. The saturated emission from the edges is parasitic emission through the etched region surrounding the lattice. 

% \begin{figure*}
% \centering
% \includegraphics[width=0.5\textwidth]{rsp_S0_pbands.png}
% \caption{(a), (c) Measured real space total intensity $S_0$ under resonant $p$ band excitation at the $K$ (a) and $K'$ (c) points with a horizontally polarized pump. (b), (d) Corresponding results for a vertically polarized pump.}
% \end{figure*} 

\end{document}